\documentclass{article}

\usepackage{arxiv}
\usepackage{amsmath}
\usepackage[utf8]{inputenc} % allow utf-8 input
\usepackage[T1]{fontenc}    % use 8-bit T1 fonts
\usepackage{hyperref}       % hyperlinks
\usepackage{url}            % simple URL typesetting
\usepackage{booktabs}       % professional-quality tables
\usepackage{amsfonts}       % blackboard math symbols
\usepackage{nicefrac}       % compact symbols for 1/2, etc.
\usepackage{microtype}      % microtypography
\usepackage{lipsum}
\usepackage{graphicx}
\usepackage[group-separator={,},group-minimum-digits=4]{siunitx}
\usepackage[percent]{overpic}
\usepackage{natbib}
\usepackage{booktabs}
\usepackage{threeparttable}
\usepackage{siunitx}
\newcommand{\simtabnote}{\textbf{Bias} and \textbf{\% Bias} are, respectively, the empirical bias and relative bias of the PTE estimators $\widehat{R}_S$; \textbf{ESE} is the empirical standard error of $\widehat{R}_S$; \textbf{ASE} is the average of the standard error estimator $\widehat{\textrm{SE}}(\widehat{R}_S)$; \textbf{CP-Q} and \textbf{CP-N} are the empirical coverage probabilities of the 95\% confidence intervals based on the normal approximation and quantiles, respectively, for $R_S$; \textbf{RE} is the empirical relative efficiency to the Gold Standard (for the same PTE estimator). All entries are based on \num{1000} replicates. Abbreviations for the methods: inverse probability weighting (IPW) and semiparametric maximum likelihood estimation (SMLE).}
\def\supone{^{(1)}}
\def\supzero{^{(0)}}

\def\supg{^{(z)}}
\def\bbeta{\pmb{\beta}}

\def\E{\textrm{E}}
\def\I{\textrm{I}}

\title{Correcting for Missing Data When Evaluating Surrogate Markers in a Clinical Trial}

\author{ 
\href{https://orcid.org/0000-0001-5380-2427}{\includegraphics[scale=0.06]{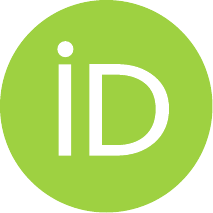}\hspace{1mm}Sarah C.~Lotspeich} \\
Department of Statistical Sciences, Wake Forest University, Winston-Salem, North Carolina, U.S.A. \\
\And 
P.D. Anh.~Nguyen \\
Department of Statistical Sciences, Wake Forest University, Winston-Salem, North Carolina, U.S.A. \\
\And 
\href{https://orcid.org/0000-0002-7288-1009}{\includegraphics[scale=0.06]{orcid.pdf}\hspace{1mm}Layla ~Parast} \\
Department of Statistics and Data Sciences, University of Texas at Austin, Austin, Texas, U.S.A. \\
\texttt{parast@austin.utexas.edu}
}

\begin{document}
\maketitle

\begin{abstract}
Evaluating treatment effects is critical in clinical trials but sometimes involves lengthy, invasive, or costly follow-up procedures. In these cases, surrogate markers, which provide intermediate measures of the long-term treatment effect, allow clinicians to obtain results faster and more efficiently than would have otherwise been possible. Prior to adoption, it is vital that the utility of surrogate markers (i.e., their ability to capture the treatment effect on the primary outcome) is statistically validated. Many frameworks for evaluating surrogate markers have been proposed, but they do not account for missing data. Instead, they rely on complete cases (the subset of patients without missing data), which can be inefficient and biased. To improve on this, we propose methods to accommodate missing data in nonparametric and parametric surrogate evaluation via inverse probability weighting (IPW) and semiparametric maximum likelihood estimation (SMLE). Through simulation studies, we demonstrate that the proposed methods remain unbiased under a broader range of missing data mechanisms than complete case analysis and can help retain the statistical precision of the full trial. We illustrate their practical utility through an application to a diabetes clinical trial. Moreover, our missing data corrections have complementary strengths with respect to computational ease, robustness, and statistical efficiency. All methods are implemented in the \textit{MissSurrogate} R package.
\end{abstract}


\section{Motivation}
When evaluating a treatment effect requires lengthy, invasive, or costly follow-up, surrogate markers provide an opportunity to assess efficacy more rapidly and with fewer resources than would otherwise be possible. A surrogate marker is a measurement intended to \textit{replace} a primary clinical outcome when evaluating a treatment, thereby enabling earlier or more efficient decision-making, and if effective, getting treatments to patients in need sooner. In the United States, the Food and Drug Administration (FDA) allows for early approval of a drug based on evidence of a beneficial effect on a surrogate marker via the Accelerated Approval Program. \citep{FDA_program,FDA_approved} For example, a clinical trial among individuals with Type $1$ diabetes is currently ongoing to examine the effect of Diamyd, a drug for the preservation of endogenous insulin production, on C-peptide levels, after the FDA agreed to consider C-peptide as a surrogate marker that would meet the requirements for the accelerated approval pathway.\citep{latres2024evidence,nowak2022intralymphatic,taylor2023c,diamyd2024fda}

Advances in the development and evaluation of surrogate markers have played a central role in accelerating approval for treatments in many disease areas, including metabolic disorders, cancer, Alzheimer's disease, and infectious diseases. However, despite their practical appeal, surrogate markers are notoriously difficult to evaluate. Even when the surrogate is strongly associated with the primary outcome, the treatment effect on the surrogate may not accurately reflect the treatment effect on the outcome itself, potentially leading to misleading conclusions. \citep{fleming1994, katz2004} Several statistical frameworks has been proposed to formalize surrogate evaluation, including the proportion of treatment effect explained framework, principal stratification, and meta-analytic approaches. \citep{elliott2023surrogate,prentice1989surrogate,freedman1992statistical,gilbert2008evaluating,burzykowski2005evaluation} However, these existing frameworks generally assume there there is no missingness. When these methods are used in practice, missing data are typically dropped such that the methods are applied to complete data only.\citep{inker2016early, sprenger2020association, Han2022, Ruilope2022, Tuzi2024, Zhuang2024}  In real-world trials, missingness frequently affects the surrogate, the clinical outcome, or both (e.g., due to dropout, failed assays, or missed visits). Yet, many surrogate evaluation methods rely on complete-case analyses, conditioning only on those individuals with both measurements, which can compromise validity and efficiency (statistical precision). 

Although some prior work has addressed surrogate marker evaluation in the presence of missing data, existing approaches remain limited. For example, Elliott et al. \citep{elliott2013accommodating} proposed a fully Bayesian approach that extends the principal stratification framework to accommodate missingness in a binary primary outcome, assuming a fully observed binary surrogate, and relying on latent ignorability or missing completely at random assumptions. Other available methods come from the broader mediation analysis literature, which is closely related but differs from surrogate evaluation in structure and assumptions. These methods are also almost entirely fully parametric. For example, Zhang and Wang \citep{zhang2013methods} develop fully parametric approaches for mediation with missing mediators or outcomes, leveraging auxiliary variables, and examining multiple imputation via multivariate normal modeling and a two-stage maximum likelihood approach. Li and Zhou \citep{li2017identifiability} investigate causal mediation with missing outcomes using estimating-equation and inverse weighting approaches, but identification relies on a parametric outcome model, a parametric mediator distribution, and the existence of a nonresponse/missingness instrument. Finally, Pesigan and Cheung \citep{pesigan2024monte} consider Monte Carlo confidence intervals for the indirect effect (in mediation analysis) with missing data, again within a fully parametric joint distributional framework. Taken together, the literature on missingness in surrogate evaluation is extremely limited, and the more extensive mediation-based work depends on restrictive fully parametric joint models for the surrogate, outcome, and missingness processes. This creates a clear need for flexible, semiparametric or model-robust approaches to surrogate evaluation that can accommodate missingness without relying on strong distributional assumptions. 

Outside of the surrogate evaluation setting, missing data methods can more generally be categorized as either design-based (e.g., relying on probabilities of being non-missing) or model-based (e.g., assuming additional mechanisms to describe the missingness), and these two classes boast different relative strengths. Inverse probability weighting (IPW) is a common design-based approach, wherein the subset of observations with non-missing values are reweighted to better represent the entire original sample. These weights can be robustly estimated (e.g., using nonparametric probabilities), but IPW is generally inefficient, since observations with missing values are still subsequently excluded (as with complete case). Maximum likelihood estimation and multiple imputation are common model-based methods, both of which assume additional models to relate the variable(s) with missing data back to the observed ones. Model-based estimators tend to be more efficient, because they use as much of the available data as possible (e.g., by filling in the missing information and keeping all observations). Still, since they require assuming an extra model, parametric model-based methods can be less robust than design-based ones. Semiparametric model-based approaches often provide a nice balance. 

Of course, the current practice of relying solely on a complete-case analysis when evaluating a surrogate  can yield substantial losses in statistical power and, depending on the mechanism of missingness, may introduce bias. In this paper, our goal is to develop methods that enable robust and efficient evaluation of surrogate markers in the presence of missing data, without relying on overly restrictive assumptions that are difficult to justify in applied settings. We first describe existing parametric and nonparametric approaches to evaluate a surrogate marker with \emph{complete} data, and then propose new methods that correct these approaches in the presences of missing surrogate marker data. We examine the performance of these methods through an extensive simulation study before applying  them to a real diabetes clinical trial dataset.  

\section{Surrogate Evaluation}
\subsection{Notation and Setting}
Let $Y$ denote the primary outcome, $S$ denote the surrogate marker, and $Z$ denote the binary treatment indicator where treatment is randomized and $Z \in \{0,1\}$ (i.e., treatment vs. control). We assume both $S$ and $Y$ are continuous. We use potential outcomes notation where each patient has a potential $\{Y \supone, Y \supzero, S \supone, S \supzero\}$ such that $Y \supg$ is the outcome when $Z=z$ and $S \supg$ is the surrogate when $Z=z$. The observed data consist of $\{Y_{zi}, S_{zi}\}$ for patient $i$ in treatment group $z$. Let $n_z$ denote the number of individuals in treatment group $z$.

We focus on evaluating a surrogate marker within the  proportion of treatment effect explained framework.\citep{freedman1992statistical,wang2002measure,parast2015robust} This framework involves estimating the proportion of treatment effect on the primary outcome that is explained by the treatment effect on the surrogate marker, often shortened to the \textit{proportion of treatment effect explained}. This quantity is defined by considering a contrast between the overall treatment effect on $Y$, 
\begin{align*}
\Delta &= E(Y \supone - Y \supzero) = E(Y \supone) - E(Y \supzero),
\end{align*}
and the residual treatment effect on $Y$, 
\begin{align*}
\Delta_S &= \int E(Y \supone - Y \supzero | S \supone = S \supzero = s) dF_{S^{(0)}}(s)    
\end{align*}
where $F_{S^{(0)}}(s) = \Pr(S \leq s|Z=0)$ is the cumulative distribution function (CDF) of $S \supzero$. This latter quantity, $\Delta_S$, measures the ``leftover" treatment effect on $Y$ after accounting for the treatment effect on the surrogate marker $S$. If a surrogate completely captures the treatment effect on $Y$, we would expect $\Delta_S$ to be zero. In contrast, if a surrogate captures none of the treatment effect on $Y$, we would expect $\Delta_S=\Delta$. Importantly, $\Delta_S$ as defined above is not identifiable, as $S\supzero$ and $S\supone$ are never observed for the same individual. Thus, the following identifiability assumption is required: 
\begin{enumerate}
\item[] (A$1$) ~~~$Y \supone \perp S \supzero | S \supone$ and $Y \supzero \perp S \supone | S \supzero$,
\end{enumerate}
such that under (A$1$), we may re-express $\Delta_S$ as $$\Delta_S = \int E(Y \supone | S \supone = s) dF_{S^{(0)}}(s) - E(Y\supzero).$$ Thus, the proportion of treatment effect explained (PTE) by the surrogate is defined as: 
$$R_S = \frac{\Delta - \Delta_S}{\Delta} = 1-\frac{\Delta_S}{\Delta},$$ where $R_S=1$ indicates a perfect surrogate and $R_S=0$ indicates a useless surrogate. With complete (non-missing) data, methods have been proposed to estimate $R_S$ both parametrically and nonparametrically.\citep{freedman1992statistical,wang2002measure,parast2015robust} 

\subsection{Parametric estimation} \label{par}
Parametric estimation of $R_S$, as proposed in Wang and Taylor (2002),\citep{wang2002measure} proceeds as follows. First, a parametric model describing the relationships between $Y$, $S$ and $Z$ is specified. For example, one may consider the following linear regression specification which includes some model flexibility via the interaction effect between $S$ and $Z$: 

\begin{equation}
    E(Y\mid Z,S) = \beta_0 + \beta_1Z +\beta_2S + \beta_3SZ  \label{parametric}
\end{equation}
and define the conditional means of the surrogate marker in each treatment group as 
\begin{equation}
     \alpha_1 = E(S\mid Z=1) \quad \text{ and } \quad \alpha_0 = E(S\mid Z=0). \label{p3}
\end{equation}
It can be shown that, when model (\ref{parametric}) holds, the overall treatment effect on $Y$ is 
\begin{eqnarray}
    ~~\Delta  &=&\beta_1 + (\beta_2 + \beta_3)\alpha_1  - \beta_2\alpha_0,  \label{Delta_param} 
\end{eqnarray}
and the residual treatment effect on $Y$ is 
\begin{eqnarray}
    \Delta_S
          &=&   \beta_1 + \beta_3\alpha_0. \label{DeltaS_param}
\end{eqnarray}
Thus, proportion of treatment effect explained by the surrogate is 
\begin{eqnarray}
R_S &= \frac{(\beta_2 + \beta_3)(\alpha_1  -\alpha_0)}{\beta_1 + (\beta_2 + \beta_3)\alpha_1  - \beta_2\alpha_0}. \label{RS_param}
\end{eqnarray}
Therefore, one may estimate $R_S$ by (i) fitting model  (\ref{parametric}) and plugging in the corresponding estimates of $\bbeta = (\beta_0, \beta_1, \beta_2, \beta_3)^\top$ and (ii) estimating $\alpha_0$ and $\alpha_1$ as just the simple means of the surrogate within each treatment group. It has been previously shown that this estimate works well, in terms of minimal bias, when model  (\ref{parametric}) holds, but can be biased, or more specifically ill-defined, when this model does not hold.\citep{parast2015robust}  

\subsection{Nonparametric estimation} \label{nonpar}
As an alternative to parametric estimation, one can consider a fully nonparametric estimation approach as proposed in Parast et al. (2016).\citep{parast2015robust} The overall treatment effect on $Y$, $\Delta$, can be estimated as simply the difference in the treatment-specific means of $Y$: 
\begin{align}
\widehat{\Delta} &= \frac{1}{n_1}\sum_{i=1}^{n_1}Y_{1i} - \frac{1}{n_0}\sum_{i=1}^{n_0}Y_{0i}, \label{Delta_np}
\end{align}
where $Y_{0i}$ ($i \in \{1, \dots, n_0\}$) and $Y_{1i}$ ($i \in \{1, \dots, n_1\}$) are the primary outcomes for the $i$th patients with $Z = 0$ and $Z = 1$, respectively. The residual treatment effect on $Y$, $\Delta_S$, can be estimated as:
\begin{equation} \widehat{\Delta}_S = \int \widehat{\mu}_1(s) d\widehat{F}_{S^{(0)}}(s) -  \frac{1}{n_0}\sum_{i=1}^{n_0}Y_{0i} \label{DeltaS_nonparam}
\end{equation}
where $\widehat{F}_{S^{(0)}}(s)$ is the estimated empirical CDF of $S^{(0)}$ and 
 $\widehat{\mu}_1(s)$ is the Nadaraya-Watson kernel-based conditional mean estimator of $\mu_1(s)  =E(Y\supone|S\supone)$. That is, 
 $$\widehat{\mu}_1(s)=\frac{\sum_{i=1}^{n_1}K_{h}(S_{1i}-s)Y_{1i}}{\sum_{i=1}^{n_1}K_{h}(S_{1i}-s)}, $$
where $K_h(x)=K(x/h)/h$ is made up of a smooth symmetric density function with finite support $K(x)$ and an appropriate smoothing bandwidth $h$. The bandwidth $h$ may be data dependent. Thus, \eqref{DeltaS_nonparam} can be expressed as:
\begin{equation*} \widehat{\Delta}_S =\frac{1}{n_0}\sum_{i=1}^{n_0} \left \{  \frac{\sum_{j=1}^{n_1}K_{{h}}(S_{1j}-S_{0i})Y_{1j}}{\sum_{j=1}^{n_1}K_{{h}}(S_{1j}-S_{0i})} \right \}- \frac{1}{n_0}\sum_{i=1}^{n_0}Y_{0i}.%\label{eq:nadaraya-watson}
\end{equation*}
The nonparametric estimate of $R_S$ is then $\widehat{R}_{S} = 1- \widehat{\Delta}_{S}/\widehat{\Delta}$. Notably, because this estimator relies on kernel smoothing over the surrogate marker $S$ through $K_h(x)$, it requires the following additional assumption:

\begin{enumerate}
\item[] (A$2$) ~~~$S$ is a continuous random variable with finite support on an interval $[a,b]$, and the distributions of $S^{(0)}$ and $S^{(1)}$ share the same support.
\end{enumerate}

Assumption (A$2$) can be partially assessed in practice by comparing the empirical distributions of $S\supzero$ and $S\supone$ in the observed data. This assumption is needed because the conditional mean function $\widehat{\mu}_{1}(s)$ is defined only on the observed support of $S\supone$. Consequently, the nonparametric PTE estimation procedure breaks down if observed values of $S\supzero$ fall outside the observed support of $S\supone$. Formally, if there exists an index $i$ such that
$$S_{0i} \notin \left\{\min_{j}(S_{1j}),\, \max_{j}(S_{1j})\right\},$$
then $\widehat{\mu}_{1}(S_{0i})$ is undefined. Although one could numerically impute such values using nearest-neighbor extrapolation, this approach can perform poorly and  introduce substantial bias near the boundary. We further examine the practical implications of support mismatch, and illustrate its impact on estimation accuracy, in our simulation studies.

\section{Surrogate Evaluation with Missing Data}\label{sec:missing}

In this section, we propose methods for surrogate evaluation when there is missingness in the surrogate, $S$. Let $O$ denote an indicator of the surrogate marker being observed, meaning that $O = 1$ if $S$ is non-missing and $O = 0$ otherwise. For simplicity, assume that $O$ inherits the same subscripts as $Y$ and $S$. That is, $O_{0i}$ ($i \in \{1, \dots, n_0\}$) and $O_{1i}$ ($i \in \{1, \dots, n_1\}$) are the indicators for the $i$th patients with $Z = 0$ and $Z = 1$, respectively. %Denote the number of patients with non-missing surrogate markers in treatment group $z$ by $\tilde{n}_z = \sum_{i=1}^{n_z}O_{zi}$ ($z \in \{0, 1\}$), where $\tilde{n}_z \leq n_z$ necessarily. 

\subsection{Complete case estimation}\label{subsec:cc}

For both the parametric and nonparametric estimation procedures, a \textit{complete case analysis} would be the simplest option. In fact, some available software, like the statistical programming language R,\cite{Rlang} defaults to complete case estimation when presented with missing data. This is also the most common approach seen across the clinical trials literature.\citep{inker2016early, sprenger2020association, Han2022, Ruilope2022, Tuzi2024, Zhuang2024} The complete case analysis would involve simply deleting all patients for whom the surrogate marker measurement is missing and applying the estimation procedures described above. 

To adapt the parametric PTE estimation approach, 
the model in \eqref{parametric} is estimated using only the subset of patients with non-missing surrogate markers to obtain complete case estimates $\hat{\bbeta}^{c}$. Then, the conditional means of the surrogate marker in the two treatment groups are calculated using the same subset of patients, denoted as 
\begin{align*}
    \hat{\alpha}_0^{c} &= \frac{\sum_{i=1}^{n_0}O_{0i}S_{0i}}{\sum_{i=1}^{n_0}O_{0i}} \text{~~ and~~ } \hat{\alpha}_1^{c} = \frac{\sum_{i=1}^{n_1}O_{1i}S_{1i}}{\sum_{i=1}^{n_1}O_{1i}}. 
\end{align*}
Finally, these estimates are substituted into \eqref{Delta_param} and \eqref{DeltaS_param} to obtain the complete case parametric estimates.  To adapt the nonparametric PTE estimation approach, 
$\widehat{\Delta}$ and $\widehat{\Delta}_S$, in (\ref{Delta_np}) and (\ref{DeltaS_nonparam}), respectively, are calculated using only complete cases.  

However, by excluding patients who are missing the surrogate marker, we face potential bias first in both estimation approaches. Within the parametric approach, the estimation of the model parameters $\hat{\bbeta}^{c}$ and calculation of the conditional means $\hat{\alpha}_0^{c}$ and $\hat{\alpha}_1^{c}$ can be biased by this exclusion. For the nonparametric approach, estimates of $\mu_1(s)$ and $F_{S\supzero}(s)$ based on complete cases may be biased. In the linear regression model, the complete case estimates $\hat{\bbeta}^{c}$ will be consistent for the true model coefficients $\bbeta$ only if the subset of patients with non-missing surrogate markers are representative of \textit{all} patients with respect to this relationship between $Y$, $S$, and $Z$ set forth in \eqref{parametric}. Similarly, the complete case estimates of $\mu_1(s)$ and $F_{S\supzero}(s)$ will be consistent only if this representativeness holds as well. When the surrogate marker is missing completely at random (MCAR), such that $O$ is independent of $Y, S$, and $Z$, the complete case analysis will be a consistent, yet inefficient, approach. However, if the surrogate marker is instead missing at random (MAR) (e.g., $O$ depends on $Y$), the complete case analysis may no longer be consistent and more sophisticated design or model-based corrections are needed. In simulations, we examine performance of the proposed methods in settings where $S$ is MCAR and MAR; across these scenarios, examples are provided where the complete case analysis is and is not biased.

\subsection{Design-based estimation}\label{sec:ipw}

To combat potential loss of representativeness in the complete cases, we first propose to accommodate missing surrogates using a design-based approach, specifically inverse probability weighting (IPW). Essentially, we re-weight the subset of complete cases (i.e., patients with non-missing surrogate markers) to represent what we expect to have seen across the entire sample had there been no missing values. This IPW approach can be used to adapt both the parametric and nonparametric PTE estimation approaches to handle missing surrogate markers, which we describe below.

We begin by modeling the missingness mechanism in our sample. When missingness is expected to depend on one or a few categorical variables, this model can be estimated nonparametrically. For example, if missingness depends only on the treatment assignment $Z = z$: 
\begin{align*}
    \widehat{\Pr}(O=1|Z=z) &= \frac{1}{n_z}\sum_{i=1}^{n_z}O_{zi}, 
\end{align*}
Note that though we refer to modeling the missingness, this is technically examining the probability that the surrogate is non-missing or \textit{observed} ($O=1$). %Let $w_{zi}$ denote the inverse of the (conditional) probability of patient $i$ in treatment group $z$ having a non-missing surrogate marker $S$. For example, in this case,  $w_{zi}= 1/\widehat{\Pr}(O_i=1|Z_i=z)$. 
Let $w_{1i}$ and $w_{0i}$ denote the estimated weights of the $i$th patients in the treatment and control groups, respectively (i.e., the inverse of their [conditional] probability of having a non-missing surrogate marker $S$).  These design-based weights will be used to reweight the observed data in the proposed IPW estimators below. However, if missingness depends on multiple variables, a model of some kind that is fit using the entire sample is generally needed, such as a logistic regression or generalized boosted model.

For parametric PTE estimation with IPW, we begin by obtaining estimates of $\beta_1, \beta_2$, and $\beta_3$ in \eqref{parametric} from a weighted linear regression model fit to the complete cases, weighting each patient by the estimated weight, $w_i$. Estimates of $\alpha_0$ and $\alpha_1$ are obtained as the weighted means of the surrogate within each treatment group, 
\begin{align*}
    \hat{\alpha}_0^{w} = \frac{\sum_{i=1}^{n_0}w_{0i}O_{0i}S_{0i}}{\sum_{i=1}^{n_0}w_{0i}O_{0i}}\text{~~ and ~~}\hat{\alpha}_1^{w} = \frac{\sum_{i=1}^{n_1}w_{1i}O_{1i}S_{1i}}{\sum_{i=1}^{n_1}w_{1i}O_{1i}},\nonumber
\end{align*}
which are Horvitz--Thompson estimators.\cite{HorvitzThompson1952} %Alternatively, if normality seems reasonable, $\widehat{\alpha}_0$ and $\widehat{\alpha}_1$ can be obtained by fitting a weighted linear regression model with $S$ as the outcome and $Z$ as the covariate.
Finally, these estimates can be plugged into \eqref{Delta_param}--\eqref{RS_param} to obtain the IPW parametric PTE estimator and its components.

For nonparametric PTE estimation with IPW, the estimate of the treatment effect, $\widehat{\Delta}$, remains unchanged (because this is a function of only the fully-observed primary outcomes) while the estimate of $\Delta_S$ must be adjusted to incorporate the weights. We define the IPW nonparametric estimate of $\Delta_S$ as
$$    \widehat{\Delta}^w_{S} = \frac{ \sum_{i=1}^{n_0} w_{0i}O_{0i}  \widehat{\mu}^w_{1}(S_{0i}) }{\sum_{i=1}^{n_0} w_{0i}O_{0i}}- \frac{1}{n_0}\sum_{i=1}^{n_0}Y_{0i}$$
where the updated Nadaraya-Watson kernel-based conditional mean estimator is 
    $$\widehat{\mu}^w_{1}(s)=\frac{\sum_{j=1}^{n_1}K_{{h^*}}(S_{1j}-s)w_{1j}O_{1i}Y_{1j}}{\sum_{j=1}^{n_1}K_{{h^*}}(S_{1j}-s)w_{1j}O_{1i}}$$
and $h^*$ is a bandwidth that possibly differs from $h$ (for complete data). The IPW nonparametric estimate of $R_S$ is thus $\widehat{R}^w_S = 1-\widehat{\Delta}^w_{S}/\widehat{\Delta}.$

Although IPW will yield consistent estimators under a correctly specified missingness model, it is well known that IPW can suffer from substantial finite-sample inefficiency. In particular, the reliance on estimated inverse probabilities can produce highly variable weights, leading to unstable estimates and wide confidence intervals, especially when some patients have small estimated probabilities of being observed and thus their estimated weights are very large. To address these limitations, in the next section, we additionally consider a semiparametric maximum likelihood approach, which is compatible with the parametric PTE estimation and offers the potential for improved efficiency while still accommodating the missing data mechanism. We compare these estimators within our simulation studies to assess the degree to which efficiency gains can be achieved without sacrificing robustness to missingness. We also examine the impact of misspecifying the weights model. 

\subsection{Model-based estimation}\label{smle}

Here, we propose a model-based approach to accommodate missingness, specifically a semiparametric maximum likelihood approach, within the parametric PTE estimation framework. Recall that we encounter and overcome the missing surrogates in the following two places. First, notice that estimating the linear regression in \eqref{parametric}, with $Y$ as the outcome and ($Z$, $S$) as covariates, is the crux of the parametric estimation procedure. Thus, our goal will be to estimate the $\bbeta$ parameters in the linear regression model, while leveraging fully observed information on $Z$ in order to explain missingness in the surrogate. In addition to $\bbeta$, the parametric estimation approach requires estimates of the conditional means, $\alpha_z = \E(S|Z=z)$ ($z \in \{0, 1\}$), i.e., the mean of the surrogate marker in the treatment and control groups. These conditional means are further challenged by the missingness.

Below, we propose a likelihood-based approach wherein we retain the same parametric component of Wang and Taylor's\cite{wang2002measure} likelihood  (the normal linear regression in \eqref{parametric}), while leaving the second model (the conditional distribution of $S$ given $Z$) to be estimated nonparametrically, thus resulting in a semiparametric approach.

%\\subsubsection{Overcoming missingness the linear regression model}
\subsubsection{Overview of the full-likelihood approach}
 
Any full-likelihood approach to this problem begins by considering a complete observation (i.e., one with a non-missing surrogate marker), which is assumed to be generated from a joint distribution: 
\begin{align}
    {\Pr}_{\bbeta,\sigma}(Y,S,Z) &= {\Pr}_{\bbeta,\sigma}(Y|S,Z)\Pr(S|Z)\Pr(Z),\label{cd_joint}
\end{align}
where %$\Pr(R|Y,S,Z)$ describes the \textit{missingness mechanism}, 
${\Pr}_{\bbeta,\sigma}(Y|S,Z)$ corresponds to the linear regression model in \eqref{parametric}, $\Pr(S|Z)$ models the conditional distributions of the surrogate marker in the treatment groups, and $\Pr(Z)$ is dictated by the treatment group assignment. Since $Z$ is fully observed, $\Pr(Z)$ will drop out of our calculations in pursuit of $\bbeta$, so we do not need to model it. 

When the surrogate marker is fully observed, the parameters $\bbeta$ for ${\Pr}_{\bbeta, \sigma}(Y|S,Z)$ can be estimated using a stand-alone linear regression model. %, as in Section~\ref{par}. 
However, when the surrogate marker is missing for some patients, we cannot estimate this model without also accounting for all other terms in \eqref{cd_joint} that involve $S$. That is, we need to model both ${\Pr}_{\bbeta, \sigma}(Y|S,Z)$ and $\Pr(S|Z)$.

Of course, in the parametric PTE estimation approach, we already have a model specified for ${\Pr}_{\bbeta, \sigma}(Y|S,Z)$ in (\ref{parametric}). To avoid making another distribution assumption, %i.e., an assumption about $\Pr(S|Z)$,
we choose to model $\Pr(S|Z)$ nonparametrically. Denote the numbers of unique observed values of the surrogate marker in the control and treatment groups as $m_0$ and $m_1$, respectively. Then, these probabilities can be  defined as $p_{kz} = \Pr(S=s_k|Z=z)$ ($k \in \{1, \dots, m_z\}$, $z \in \{0,1\}$). Again, if the surrogate marker were fully observed, we could use empirical probabilities: $\hat{p}_{kz} =$ $\sum_{i=1}^{n}\I\left(S_i = s_k \text{ and } Z_i = z\right)/\sum_{i=1}^{n}\I\left(Z_i = z\right)$, where $\I(\cdot)$ is the standard indicator function. Instead, we need to further account for the missingness in $S$ as we estimate the $p_{kz}$. %(We should also note that we need this distribution to estimate the conditional means of the $S$ in the next stage.) \\

For ease of notation, suppose that the probabilities are organized into vectors for the control and treatment groups: 
\begin{align*}
    \pmb{p}_0 = \begin{bmatrix}
        p_{10} \\
        \vdots \\
        p_{m_{00}}
    \end{bmatrix} \text{ and }
    \pmb{p}_1 = \begin{bmatrix}
        p_{11} \\
        \vdots \\
        p_{m_{11}}
    \end{bmatrix},
\end{align*}
respectively. Also, note that these probabilities are constrained such that $\sum_{k=1}^{m_z}p_{kz} = 1$ ($z \in \{0, 1\}$), following the axioms of probability. Conveniently, we can derive the log-likelihood function from \eqref{cd_joint}, incorporating $\pmb{p}_0$ and $\pmb{p}_1$, which can be maximized to estimate  ($\bbeta$, $\sigma$, $\pmb{p}_0$, $\pmb{p}_1$) simultaneously. The following expectation-maximization (EM) algorithm \citep{Dempster1977} helps to overcome the missingness in $S$ as we do so. 

\subsubsection{EM algorithm}

We begin by defining the \textit{complete-data log-likelihood}, i.e., the log-likelihood for ($\bbeta$, $\sigma$, $\pmb{p}_0$, $\pmb{p}_1$) if $S$ was not missing for anyone: 
\begin{align}
%&\ell(\bbeta, \sigma, \pmb{p}) \nonumber %&= \sum_{i=1}^{n}\sum_{z=0}^{1}\sum_{k=1}^{m}\I\left(S_i = s_k\right)\log\left\{{\Pr}_{\bbeta, \sigma}(Y=Y_i,S=s_k,Z=z)\right\} \nonumber \\
\ell(\bbeta, \sigma, \pmb{p}_0, \pmb{p}_1) \propto &\sum_{i=1}^{n}\sum_{k=1}^{m_{Z_i}}\I\left(S_i=s_k\right)\left[\log\left\{{\Pr}_{\bbeta, \sigma}(Y_i|s_k,Z_i)\right\} + \log\left(p_{kZ_i}\right)\right].\label{cd_loglik}
% &\sum_{i=1}^{n}\sum_{z=0}^{1}\I(Z_i=z)\sum_{k=1}^{m_z}\I\left(S_i=s_k\right)\left[\log\left\{{\Pr}_{\bbeta, \sigma}(Y_i|s_k,z)\right\} + \log\left(p_{kz}\right)\right].\label{cd_loglik}
%\sum_{i=1}^{n}\sum_{z=0}^{1}\I(Z_i=z)\sum_{k=1}^{m_z}\I\left(S_i = s_k\right)\left[\log\left\{{\Pr}_{\bbeta, \sigma}(Y=Y_i|S=s_k,Z=z)\right\} + \log\left(p_{kz}\right)\right].\label{cd_loglik}
\end{align}
While there are certainly other ways to define this log-likelihood, we preferred \eqref{cd_loglik} because the missingness in the surrogate marker has been isolated to %the indicator function, 
$\I\left(S_i = s_k\right)$. To find the semiparametric maximum likelihood estimators (SMLEs) from \eqref{cd_loglik}, we begin with noninformative initial values $\hat{\bbeta}^{(0)} = \pmb{0}$, $\hat{\sigma}^{(0)} = 0.1$, and $\hat{p}^{(0)}_{kz} = 1/m_z$ ($k = 1, \dots, m_z$, $z\in\{0,1\}$). Then, we iterate between the following E- and M-steps, updating these parameter values until they converge (within a $0.001$ tolerance).

In the \textit{E-step}, we compute conditional expectations of the missing indicators $\I\left(S_i = s_k\right)$ for each patient $i$ ($i \in \{1, \dots, n\}$) and observed surrogate value $k$ ($k \in \{1, \dots, m_z\}$) given their other fully-observed information ($Y_i, Z_i$), which is equal to the following probability: %$\hat{\phi}_{ki}(\bbeta, \sigma, \pmb{p}) $
\begin{align}
    \hat{\phi}_{ki}(\bbeta, \sigma, \pmb{p}_0, \pmb{p}_1) 
    &= %\E_{\bbeta, \pmb{p}}\left\{\I\left(S_i = s_k\right)|Y_i,Z_i \right\} = 
    {\Pr}_{\bbeta, \sigma,\phantom{j}\pmb{p}_0,\phantom{j} \pmb{p}_1}\left(S_i = s_k|Y = Y_i, Z= Z_i \right) \nonumber \\
    &= \frac{{\Pr}_{\bbeta, \sigma}(Y=Y_i|S=s_k,Z=Z_i)p_{kZ_i}}{\sum_{k'=1}^{m_{Z_i}}{\Pr}_{\bbeta, \sigma}(Y=Y_i|S=S_{k'},Z=Z_i)p_{k'Z_i}}. \label{e_step}
\end{align} 
In each iteration $t$, we evaluate \eqref{e_step} at the previous iteration's estimated parameter values $(\hat{\bbeta}^{(t-1)}, \hat{\sigma}^{(t-1)}, \hat{\pmb{p}}_0^{(t-1)}, \hat{\pmb{p}}_1^{(t-1)})$. For ease of notation, let $\hat{\phi}_{ki}^{(t-1)}$ denote $\hat{\phi}_{ki}(\hat{\bbeta}^{(t-1)}, \hat{\sigma}^{(t-1)}, \hat{\pmb{p}}_0^{(t-1)}, \hat{\pmb{p}}_1^{(t-1)})$. %, i.e., the conditional expectation evaluated at iteration $t$'s parameters, going forward.

In the \textit{M-step}, the missing indicators $\I\left(S_i = s_k\right)$ in \eqref{cd_loglik} are replaced with conditional expectations from the E-step and $(\hat{\bbeta}^{(t)}, \hat{\sigma}^{(t)}, \hat{\pmb{p}}^{(t)})$ are updated to maximize the resulting objective function: %Specifically, in the $t$th iteration, we use conditional expectations $\hat{\phi}_{ki}^{(t-1)}$ based on the previous iteration's parameters in place of the missing indicators $\I\left(S_i = s_k\right)$ and update the parameter estimates $(\hat{\bbeta}^{(t)}, \hat{\sigma}^{(t)}, \hat{\pmb{p}}^{(t)})$ to maximize 
\begin{align}
%\sum_{i=1}^{n}\sum_{z=0}^{1}\I(Z_i=z)\sum_{k=1}^{m}\hat{\phi}_{ki}^{(t-1)}\left[\log\left\{{\Pr}_{\bbeta, \sigma}(Y=Y_i|S=s_k,Z=z)\right\} + \log\left(p_{kz}\right)\right], \label{m_step}
\sum_{i=1}^{n}\sum_{k=1}^{m_{Z_i}}\hat{\phi}_{ki}^{(t-1)}\left[\log\left\{{\Pr}_{\bbeta, \sigma}(Y_i|s_k,Z_i)\right\} + \log\left(p_{kZ_i}\right)\right], \label{m_step}
\end{align}
where $\hat{\phi}_{ki}^{(t-1)} = \I(S_i = s_k)$ if patient $i$ had a non-missing surrogate marker and follows \eqref{e_step} otherwise. %For the computational implementation, it may be helpful to note that \eqref{m_step} can be rewritten with separate sums among patients in the treatment and control groups. 
%\subsubsection{Semiparametric maximum likelihood estimators}
Conveniently, $(\hat{\bbeta}^{(t)}, \hat{\sigma}^{(t)})$ can be obtained using standard software for a weighted linear regression model (with $\hat{\phi}_{ki}^{(t-1)}$ as the weights), and the updated probabilities have a straightforward analytical form: 
\begin{align}
    \hat{p}_{kz}^{(t)} &= \frac{\sum_{i=1}^{n}\I(Z_i=z)\hat{\phi}_{ki}^{(t-1)}}{\sum_{k=1}^{m_z}\sum_{i=1}^{n}\I(Z_i=z)\hat{\phi}_{ki}^{(t-1)}}. %= \frac{\sum_{i:Z_i=1}\hat{\phi}_{ki}^{(t-1)}}{\sum_{k=1}^{m}\sum_{i:Z_i=1}\hat{\phi}_{ki}^{(t-1)}}. 
    \label{m_step_p}
\end{align}
Notice that the updates in \eqref{m_step_p} continue to respect the constraint: $\sum_{k=1}^{m_z}\hat{p}_{kz}^{(t)} = 1$.  Importantly, even when the surrogate marker is continuous and all observed values are unique, this overall formulation corresponds to the nonparametric maximum likelihood estimator of 
$Pr(S|Z)$, which places probability mass on the observed support points. The benefit of this approach is the resulting ability to accommodate missing surrogates without imposing a parametric model for the distribution of $S$, e.g., by placing an additional normal linear regression on it.  \\ 

\subsubsection{Final Estimation}
The SMLE approach implemented via the EM algorithm results in estimates of 
$(\bbeta, \sigma)$ and $p_{kz}$ for all $k = 1, \ldots, m_z$ and $z \in \{0,1\}$. 
Recall that the quantities $\Delta$, $\Delta_S$, and $R_S$ depend on $\bbeta$ as well as 
the conditional means $\alpha_0$ and $\alpha_1$. Thus, although we now have an estimate 
of $\bbeta$ directly from maximizing the likelihood, we still require estimates of $\alpha_0$ and 
$\alpha_1$.

With the estimated $p_{kz}$, these quantities are straightforward to obtain since:
\begin{eqnarray*}
\alpha_0 &=& \E(S|Z=0) \\
&=& \sum_{k=1}^{m_0} s_k \Pr(S=s_k|Z=0)\\
&=& \sum_{k=1}^{m_0} s_k p_{k0}, 
\end{eqnarray*}
and similarly for $\alpha_1$, and thus we can estimate these quantities using the SMLE as: 
\begin{align}
\hat{\alpha}_0^{l} = \sum_{k=1}^{m_0}s_k\hat{p}_{k0} \text{ ~and~ } \hat{\alpha}_1^{l} = \sum_{k=1}^{m_1}s_k\hat{p}_{k1}. \label{mle_alpha}
\end{align}
With these components, we then obtain $\widehat{\Delta}^{l}$ and $\widehat{\Delta}_S^{l}$  by plugging the SMLEs $(\widehat{\bbeta}^{l}, \hat{\alpha}_0^{l}, \hat{\alpha}_1^{l})$ into \eqref{Delta_param} and \eqref{DeltaS_param}, respectively, and then calculate $\widehat{R}_S^{l} = 1 - \widehat{\Delta}_S^{l}/\widehat{\Delta}^{l}$. The resulting estimator is fully likelihood-based and accommodates missing surrogate data through a nonparametric model for $\Pr(S \mid Z)$.

\subsection{Choosing between design- and model-based estimation}

As in other settings, the design- and model-based missing data corrections offer complementary strengths. The selection of which correction to adopt in practice could depend on a number of theoretical and logistical features. In terms of robustness, the two corrections are fairly similar, since the SMLE is not as strict as a fully-parametric approach would be (e.g., one that assumes an additional linear regression model for the conditional distribution of $S$ given $Z$). Broadly, decision-making can come down to (i) whether one has the computational and statistical capacity to support a data-driven correction like the SMLE versus (ii) how confident one is in their ability to correctly specify the missingness mechanism for IPW. 

First, the model-based SMLE approach will generally offer better statistical efficiency (i.e., narrower confidence intervals), because all patients contribute directly to the model. For the design-based IPW approach, patients with missing surrogates can only lend information through the estimation of the weights, meaning that their non-missing primary outcomes $Y$ are omitted. However, computational runtimes for the SMLE are much longer than for IPW, which is essentially as fast as fitting the complete case analysis, and the SMLE could struggle under heavy missingness if there are too few values $s_k$ with which to nonparametrically estimate $\Pr(S|Z)$. 

Second, the design-based IPW approach provides flexibility in handling the missing data, since different modeling approaches (e.g., empirical probabilities or logistic regression) can be used to estimate the weights. If missingness depended on more than the treatment $Z$ or primary outcome $Y$, it would also be very straightforward to accommodate any other patient characteristics (like age) as additional covariates in the weights model. Meanwhile, the SMLE would require smoothing (e.g., with B-spline sieves or a kernel smoother) for the nonparametric estimate of the conditional distribution to include more than just $Z$, and it would not be plausible to add very many variables to this model due to the curse of dimensionality. %Still, under simpler missingness mechanisms or when unsure how to fully specify the model, the SMLE could be beneficial. 

\subsection{Standard error estimation and confidence interval construction}\label{sec:se}

The standard errors for the estimators of the key quantities $\Delta$, $\Delta_S$, and $R_S$ can be estimated using a bootstrap resampling approach. Conveniently, this approach can be used to calculate standard errors for either the parametric or nonparametric PTE estimator \emph{and} using any of the missing data methods outlined. Specifically, we repeat the following two steps $D = 500$ times, subsequently using the results to approximate the distributions of the estimators for $\Delta$, $\Delta_S$, and $R_S$. For each iteration $d$, % ($d \in \{1, \dots, D\}$, 
we proceed as follows. 
\begin{enumerate}
      \item \textit{Bootstrap resampling:} Randomly resample with replacement $n_0$ patients from the control group and $n_1$ patients from the treatment group. Stratifying the resampling in this way ensures that the treatment rate $\Pr(Z = 1)$ is retained in all iterations. 
    \item \textit{PTE Re-estimation:} Using the resampled patients, re-estimate $\Delta$, $\Delta_S$, and $R_S$ using either PTE estimator and any missing data method. Denote estimates from this iteration by $\widehat{\Delta}^{(d)}$, $\widehat{\Delta}_S^{(d)}$, and $\widehat{R}_S^{(d)}$. 
\end{enumerate}
At the end of this procedure, the $D$ estimates of $\Delta$, $\Delta_S$, and $R_S$ are used to (i) estimate the variance of the estimators and (ii) construct confidence intervals.

First, each estimator's variance can be estimated with the sample variance of the $D$ values obtained after bootstrap resampling, denoted by $\widehat{\textrm{V}}(\widehat{\Delta})$, $\widehat{\textrm{V}}(\widehat{\Delta}_S)$, and $\widehat{\textrm{V}}(\widehat{R}_S)$. Second, a Wald-type $95\%$ confidence interval for each quantity can be constructed based on a normal approximation using the estimated variance. Using $\widehat{R}_S$ for demonstration, this interval is defined as $\left\{\widehat{R}_S \pm 1.96 \sqrt{\widehat{\textrm{V}}(\widehat{R}_S)}\right\}$. Third, quantile-based $95\%$ confidence intervals can be built based on the central $95\%$ of the bootstrapped values. For $\widehat{R}_S$, this alternative interval is defined as $(\hat{q}_{0.025}, \hat{q}_{0.975})$, where $\hat{q}_p$ ($p \in [0, 1]$) denotes the $p$th quantile of the $D$ bootstrapped values $\widehat{R}_S^{(1)}, \dots, \widehat{R}_S^{(D)}$. As in other settings, we expect the quantile-based confidence intervals to be more robust, while the Wald-type ones will perform best in larger samples.  %The variance estimates for $\widehat{\Delta}$ and $\widehat{\Delta}_S$ and corresponding confidence intervals are obtained following the same procedures. 

\section{Simulation Studies}\label{sims}

%\textcolor{blue}{[From Layla: we need to re-run these to use bootstrapping for all, and also use the new IPW NP estimator. And also the limit seems to be 6 total tables and figures :-/]} 
Through extensive simulation studies, we examine how our proposed IPW and SMLE approaches correct for bias in the parametric or nonparametric PTE estimators due to missing surrogate markers. We further confirm the validity of their standard error estimators and bootstrap confidence intervals, integral metrics when assessing a potential surrogate's strength. %Finally, we demonstrate the relative advantages of the two approaches, including the statistical efficiency and robustness of the SMLE, and the computational ease of IPW. 
These simulations were conducted in the statistical programming language R,\citep{R} and all code and data can be found %on GitHub 
at \url{https://github.com/sarahlotspeich/evaluate_missing_surrogates}.

%In comparing these approaches, we are primarily interested in (i) which ones achieve unbiased estimates of $R_S$ despite missingness in $S$ and (ii) which ones can do so mostly efficiently (i.e., with the smallest empirical variability). 

\subsection{Data Generation}\label{subsec:dgm}

%\textcolor{blue}{[From Layla - I edited this a bit, can you check to make sure you are ok with it?]}
In all of the simulations that follow, data were generated for trials of $n = 2000$ patients, with half them assigned to the treatment arm ($Z = 1$) and half to the control arm ($Z = 0$). We considered five simulation settings which differed in the missingness mechanism for $S$.  In Settings $1$ through $4$, the surrogate marker $S$ was simulated, conditionally on the treatment $Z$, from a normal distribution: 
\begin{align*}
    S|Z &\sim \begin{cases}
        \textrm{Normal}(\mu = 6, \sigma^2 = 4) & \text{ if }Z = 1 \text{, and} \\
        \textrm{Normal}(\mu = 5, \sigma^2 = 1) & \text{ if }Z = 0,
    \end{cases}
\end{align*}
and the primary outcome $Y$ was created from ($Z$, $S$) following a linear model: $Y = 2 + Z + 5S + ZS + \epsilon$, where the random errors $\epsilon$ were independently and identically distributed $\textrm{Normal}(\mu = 0, \sigma^2 = 1)$. Based on these parameters, $\Delta = 12$ and $\Delta_S=6$, such that the true $R_S=%1 - 6 / 12 = 
0.5$. Note that, throughout, the model assumed by the parametric estimation approach is correctly specified, since $Y$ is simulated in agreement with \eqref{parametric}. It has been well-established that if this model is not correctly specified, the parametric approach, even with fully observed, will be biased.\citep{parast2015robust} As our focus is on missingness, we did not focus on this already-known result of primary model misspecification.

In each setting, the indicator of being observed $O$ was generated from a Bernoulli distribution, but depending on different variables. In Setting $1$, the surrogate was \textit{missing completely at random} such that $\Pr(O = 1) = 0.65$, leading to approximately $32$--$39\%$ missing $S$ per replication. In Setting $2$, the surrogate was \textit{missing at random given $Z$} such that $\Pr(O = 1|Z) = 1/\left\{1 + \exp\left(-0.4-0.2Z\right)\right\}$, leading to approximately $34$--$41\%$ missing $S$ per replication. In Setting $3$, the surrogate was \textit{missing at random given $Y$} such that 
$\Pr(O = 1|Y) = 1/\left\{1 + \exp\left(-0.015Y\right)\right\}$, leading to approximately $35$--$41\%$ missing $S$ per replication. In Setting $4$, the surrogate was \textit{missing at random given the interaction of $Y + YZ$} such that $\Pr(O = 1|Y,Z) = 1/\left\{1 + \exp\left(-0.015Y-0.015YZ\right)\right\}$, leading to approximately $28$--$35\%$ missing $S$ per replication. Setting 4 was designed to examine the impact of misspecifying the model used for the inverse probability weights. All settings led to a moderate amount of patients with missing surrogate markers; the difference between Settings $1$--$4$ is with respect to \textit{why} the missingness occurred, rather than \textit{how much} missingness was present. 

\begin{figure}[ht]
    \centering
    \caption{The nonparametric estimator $\widehat{R}_S$, of the proportion of treatment effect explained (PTE), assumes that the distributions of the surrogate marker $S$ in the treatment and control groups overlap (left panel). In Setting $5$, we consider the setting where these distributions do not fully overlap (right panel).
    }\label{fig:overlap}
    \includegraphics[width=0.75\linewidth]{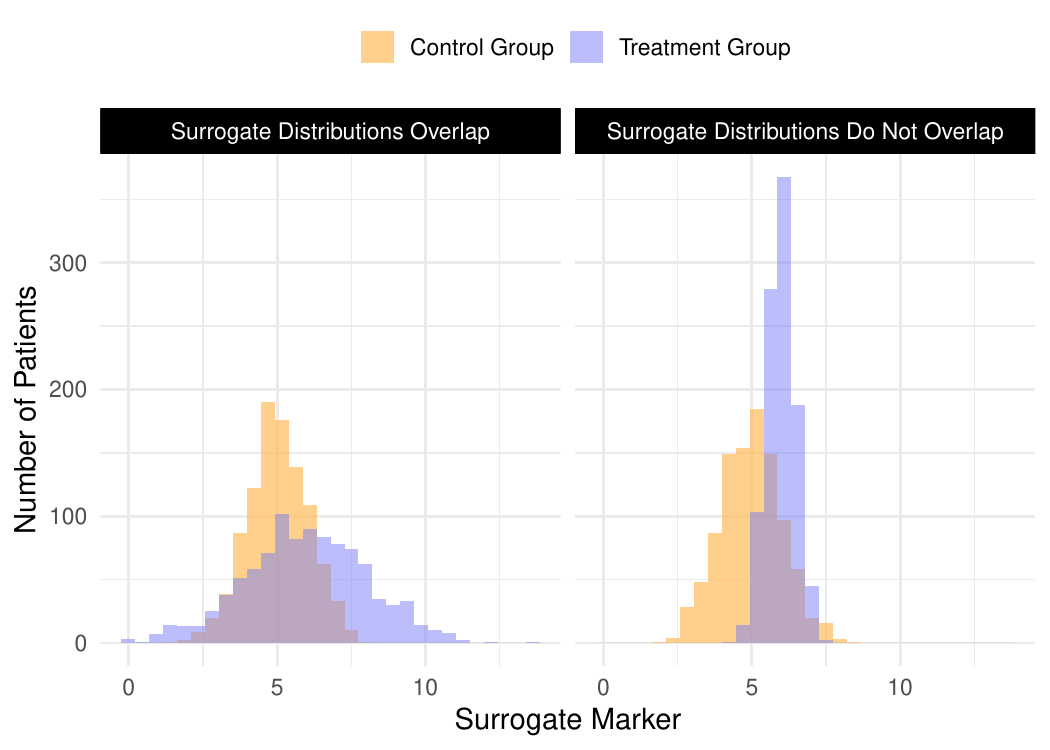}
\end{figure}

In Setting $5$, we purposefully generated data such that there was \textit{not} complete overlap in the distribution of the surrogate in the control and treatment groups; see Figure \ref{fig:overlap}. We were interested in this setting because the nonparametric PTE estimator requires the distribution of $S$ in the two groups to overlap and even though extrapolation can be used to bridge the two groups when they do not overlap, bias can persist. Thus, our aim in this setting was to explore this bias with the added challenge of surrogate marker missingness. In this setting, the surrogate marker $S$ was simulated, conditionally on the treatment $Z$, with revised parameters: 
\begin{align*}
    S|Z &\sim \begin{cases}
        \textrm{Normal}(\mu = 6, \sigma^2 = 1/4) & \text{ if }Z = 1 \text{, and} \\
        \textrm{Normal}(\mu = 5, \sigma^2 = 1) & \text{ if }Z = 0. 
    \end{cases}
\end{align*}
All other variables were simulated as described above, and missingness was generated as in Setting $3$, i.e., missing at random given $Y$. We were particularly interested in this setting because our diabetes clinical trial in which we apply these methods does not have full overlap in the surrogate distributions; we discuss this further in our data application. 

We considered seven approaches to estimate $R_S$ under these various types of missing data in the surrogate $S$, three of which are our proposed missingness correction approaches. The remaining four are comparators: the \textit{gold standard analyses} %nonparametric and parametric 
which used the full simulated surrogate marker $S$ for all $2000$ patients (i.e., no missingness), which of course would not be possible in practice, and the naive \textit{complete case analyses}, which only used the subset of patients with non-missing surrogates. There are four comparators because we have each, the gold standard and the complete case, using the nonparametric approach and the parametric approach. Throughout, standard error estimates and $95\%$ confidence intervals (based on the normal approximation and quantiles) were calculated using  bootstrap resampling with $D = 500$, as described in Section~\ref{sec:se}. 

Our three proposed missingness correction approaches were: (i) the IPW nonparametric approach, (ii) the IPW parametric approach, and (iii) the SMLE approach. Both IPW approaches estimated the probability that the surrogate was non-missing (i.e., observed) using a logistic regression model fit to all $2000$ patients, with $O$ as the outcome and either $Z$, $Y$, or both $Y$ and $Z$ (with an interaction) as the covariate(s) for Settings $1$--$3$, respectively. Of course, it is unfair to assume that one would correctly specify this covariate as the truth would be unknown in practice, which is why we examine the impact of misspecification of the IPW model in Setting $4$. In all settings, the SMLE approach was implemented as described above; no modifications were needed for different missingness mechanisms. 

%\begin{enumerate}
    %\item \textit{Gold standard, nonparametric:} Use the simulated surrogate marker $S$ for all $2000$ patients (i.e., assuming no missingness) with nonparametric $R_S$ estimation. 
    %\item \textit{Complete case, nonparametric:} Use nonparametric estimation with only  the subset of patients with non-missing surrogates and delete any patients with missing surrogates from the analysis. 
    %\item \textit{IPW, nonparametric:} Estimate the probability that the surrogate is non-missing (i.e., observed) using a logistic regression model fit to all $2000$ patients, with $O$ as the outcome and the covariates chosen based on the type of missingness (explained more below), and implement the proposed IPW nonparametric estimator. 
    %\item \textit{Gold standard, parametric:} Use the simulated surrogate marker $S$ for all $2000$ patients (i.e., assuming no missingness) with parametric $R_S$ estimation. 
    %\item \textit{Complete case, parametric:}  Use parametric estimation with only  the subset of patients with non-missing surrogates and delete any patients with missing surrogates from the analysis. 
    %\item \textit{IPW, nonparametric:} Estimate the probability that the surrogate is non-missing (i.e., observed) using a logistic regression model fit to all $2000$ patients, with $O$ as the outcome and the covariates chosen based on the type of missingness (explained more below), and implement the proposed IPW parametric estimator. 
    %\item \textit{SMLE, parametric:} Implement the proposed SMLE estimation approach.  
%\end{enumerate}
For each setting, results across $1000$ replications were summarized by the following quantities. We report empirical bias and relative bias, calculated as the mean of the difference $(\widehat{R}_S-R_S)$ and the bias divided by the true $R_S$, respectively. To confirm their validity, average standard error estimators, computed as the mean of $\widehat{\text{SE}}(\widehat{R}_S)$, were compared to the empirical standard errors, computed as the standard deviation of $\widehat{R}_S$ across replications. Similarly, the performance of 95\% confidence intervals based on the normal approximation and quantiles were measured by their empirical coverage probability. The empirical relative efficiency (RE) to the gold standard for $\widehat{R}_S$ was calculated for the complete case analysis, IPW, and SMLE as the empirical variance of the gold standard divided by the empirical variance of the alternative estimator. RE values closer to one indicate that more efficiency was recovered through how the missing surrogate markers were handled. 

\subsection{Setting 1: Missingness Completely at Random}\label{sims_mcar}

In Setting $1$, where the surrogate marker $S$ was missing completely at random (i.e., independently of other variables), all methods were empirically unbiased in estimating $R_S$ ($<0.5\%$ bias, Table~\ref{tab:mcar}). Since the subset of patients with non-missing surrogate markers was representative of the entire sample, this result was expected. In this setting, we devote our attention to confirming the validity of the proposed IPW and SMLE approaches and comparing estimators' efficiency.  The bootstrapped CIs for all methods had empirical coverage probabilities close to the nominal $95\%$. Given the larger sample size, coverage probabilities were similar for the normal approximation and quantile-based CIs, as well. 

For the gold standard and complete case analyses, the variability of the nonparametric versus parametric PTE estimators was similar. Interestingly, for the IPW approach, the nonparametric PTE estimator was slightly more efficient than the parametric one (RE $=0.72$ versus $0.68$). For the nonparametric PTE estimator, IPW was more efficient than the corresponding complete case analysis, as well (RE $=0.72$ versus $0.69$), while IPW and complete case appeared identical for the parametric PTE estimator. The SMLE was noticeably more efficient than the complete case and IPW approaches for the parametric PTE estimator (RE $=0.79$ versus $0.68$). Such efficiency is particularly important in the surrogate setting because decisions about the validity of the surrogate are often based on the lower bound of the confidence interval for $R_S$ being above some pre-defined threshold, such as $0.50$ or $0.75$. Tighter confidence intervals reduce uncertainty around this comparison and strengthen the resulting conclusions.

\subsection{Settings 2 and 3: Missingness at Random}\label{sims_mar}

In Setting 2, where the surrogate marker $S$ was missing at random based on the treatment $Z$, all methods remained empirically unbiased ($\leq 0.5\%$, top of Table~\ref{tab:mar}). The complete case analysis continued to be valid in this setting because, while the subset of patients with non-missing surrogate markers differed from the original sample with respect to the distribution of $Z$, they were still representative with respect to the relationships between $Y$, $S$, and $Z$ needed to estimate the PTE. For the nonparametric PTE estimator, IPW was only slightly more efficient than the complete case analysis (RE $=0.72$ versus $0.69$). For the parametric PTE estimator, IPW was again identical to the complete case in terms of both bias and efficiency. The SMLE offered much better efficiency than either the complete case or IPW parametric PTE estimator (RE $=0.79$ versus $0.68$); it was more efficient than either nonparametric PTE approach, as well (RE $=0.69$--$0.72$). 

In Setting $3$, where the surrogate marker $S$ was missing at random based on the primary outcome $Y$, the complete case analysis was biased estimating $R_S$ either nonparametrically or parametrically ($4\%$, bottom of Table~\ref{tab:mar}). While this bias may seem small, it impacted the coverage probabilities for the complete case analysis, which were well below the nominal $95\%$ for both estimators ($85$--$88\%$). For the nonparametric PTE estimator, this setting is where IPW gained advantage over the complete case analysis, since it remained unbiased ($<1\%$) and continued to offer $95\%$ CIs with appropriate coverage. For the parametric PTE estimator, IPW again offered low bias ($<0.5\%$) and good coverage, but the SMLE offered similarly low bias and much better efficiency (RE $= 0.75$ versus $0.65$). 

\subsection{Setting 4: Misspecifying Inverse Probability Weights}\label{sims_misspec_IPW}

\begin{figure}[ht]
    \centering
    \caption{Simulation results for $\widehat{R}_S$, the proportion of treatment effect explained (PTE) in Setting $4$, when the surrogate marker $S$ was missing at random given $Y$ and the interaction $Y \times Z$  between it and treatment group (top row) or just the primary outcome $Y$ (bottom row).}\label{fig:setting4}
    \includegraphics[width=1\linewidth]{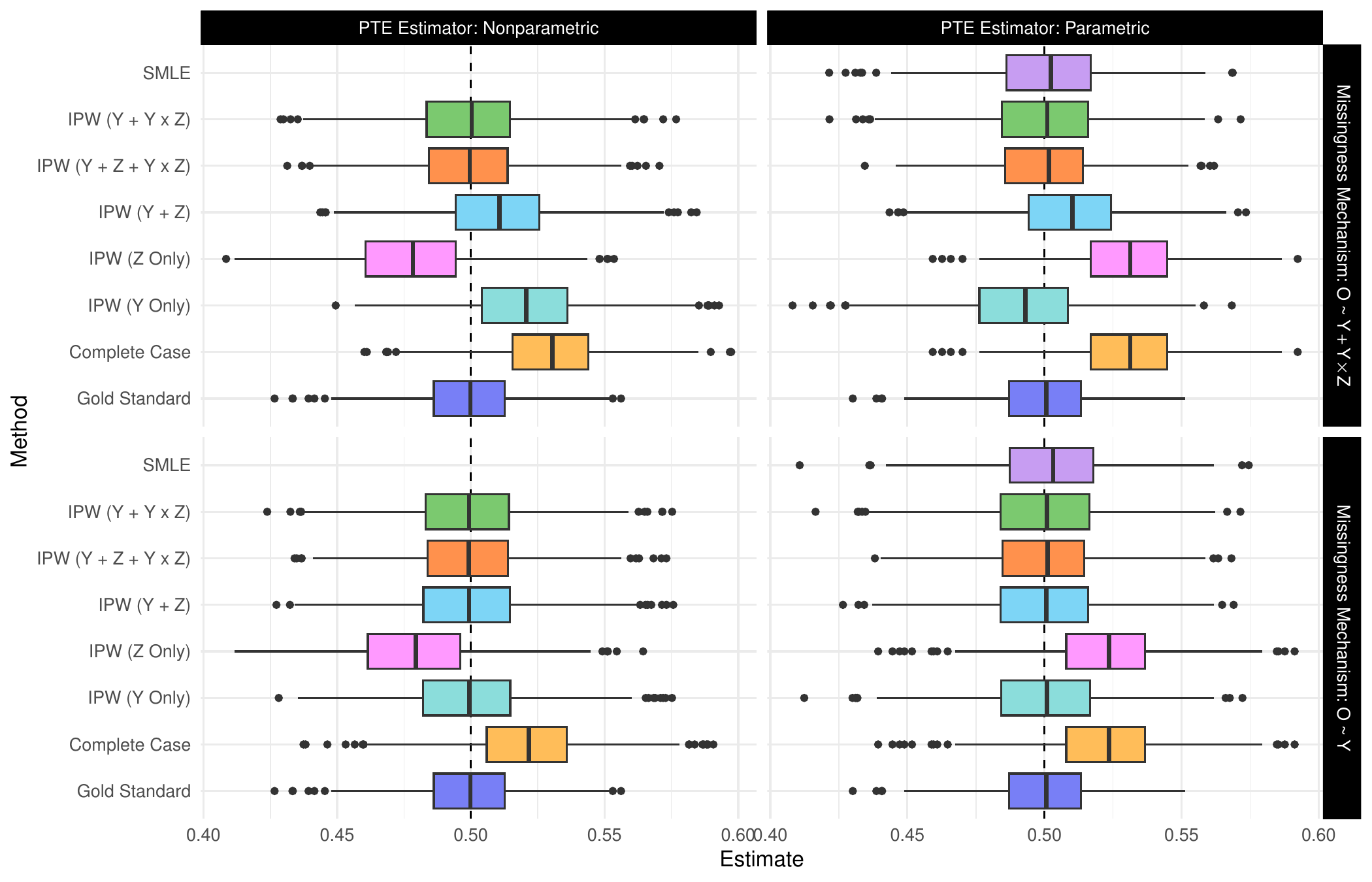}
\end{figure}

Simulation results from Setting $4$, where the surrogate marker $S$ was missing at random based on $Y$ and the interaction $Y \times Z$ between treatment group and primary outcome are shown in the top of Figure \ref{fig:setting4}. For this setting only, we show the following five versions of each (nonparametric and parametric) IPW estimator. The weights were estimated using a model that included (i) $Y$ only, (ii) $Z$ only, (iii) main effects for $Y$ and $Z$, (iv) main effects for both $Y$ and $Z$ and their interaction $YZ$, and (v) a main effect for $Y$ only and its interaction with treatment $YZ$ (the correctly-specified model). As expected, the IPW approaches to both the nonparametric and parametric PTE estimators were biased when the weights model was misspecified by excluding a key variable or term, i.e., with versions (i)--(iii). %This bias was most apparent when excluding $Y$ or $Z$ from the weights model (versions i--ii). 
When \textit{underfitting} the weights model by omitting the interaction term (version iii), the bias was present but less severe than when either $Y$ or $Z$ was not included (versions i--ii). In fact, the IPW estimator when only $Y$ or $Z$ was included in the weights model had comparable, maybe even identical, bias as the uncorrected complete case analysis. \textit{Overfitting} the weights model by including a main effect for $Z$ (version iv) led to unbiased estimates that were possibly even more precise than those using the correctly-specified model (version v). 

To continue exploring the impact of overfitting the weights model, we also revisited Setting $3$, where missingness depended only on the primary outcome $Y$ (bottom of Figure \ref{fig:setting4}). Again, we considered the five versions of both IPW estimators. The correctly specified weights model used $Y$ alone (version i), while that including $Z$ only was underfit (version ii) and those including $Y$ and $Z$ with/without their interaction (versions iii--v) include more than is necessary and are thus overfit. In general, overfitting models is expected to lead to a loss of statistical efficiency. However, since the overfit models here were used to estimate nuisance parameters, rather than the primary estimand $R_S$, it seems that the empirical efficiency when using the overfit weights models was the same if not slightly better than that with the correct one. 

Altogether, these simulations demonstrate the importance of trying to capture all relevant variables in the weights model, in their proper form, to achieve unbiased estimates with IPW. Meanwhile, a key advantage of the SMLE is that it avoids this additional level of specification by relying on the subset of patients with non-missing surrogate markers to drive the nonparametric component of the estimator. 

%\textcolor{blue}{[Sarah: Three different things you can do with missing observations, each have pros/cons. How do you pick your weights? What happens if you're wrong and in different ways (overfitting vs. underfitting)?  Overfitting the weights model loses efficiency, but remains unbiased. Underfitting the model (i.e., excluding key variables) led to bias for the IPW Nonparametric but not necessarily for the IPW parametric. (Why?? Ask Layla - Is it because the parametric is kind of recovering information because Z is still included in that final saturated model?)] From Layla: but from the picture it does seem that the parametric is biased when you leave out a key variable, right? so the message seems to be - unsurprisingly, if you mis-specify your IPW model with respect to leaving out a variable, you can be biased, that is just how it is. The advantage of the SMLE, is that we don't have to worry about that. ]}

\subsection{Setting 5: Lack of Overlap in Control and Treatment Groups' Surrogate Marker Distributions}

Simulation results for Setting $5$ are shown in Table \ref{tab:overlap}. As expected, the nonparametric estimators, even examining the gold standard, are biased. The IPW correction does not decrease this bias, since it is due to the nonparametric estimator itself failing (rather than missingess). Since $S$ here is MAR given $Y$, as in Setting $3$, we expect that some of the $20\%$ bias in the complete case estimate would be attributable to missingness. However, IPW was only slightly better at $19\%$, suggesting that the lack of overlap was a much bigger challenge in this setting than the missingness. 

\section{Data Application}
We use our proposed methods to examine a surrogate marker with missingness in the Diabetes Control and Complications Trial (DCCT).\citep{DCCT_data} The DCCT was a multicenter, randomized, clinical study designed to investigate an intensive treatment regimen directed at maintaining blood glucose concentrations close to normal levels and determine the effect on appearance or progression of early vascular complications compared to using the conventional treatment at the time of the study (one or two shots of insulin per day with daily self-monitoring of urine or blood glucose) in patients with Type $1$ diabetes.\citep{dcct1986diabetes} Trial results ultimately showed that the intensive treatment regime significantly lowered the risk of microvascular and cardiovascular complications.\citep{nathan2014diabetes}

In our analysis, the surrogate marker of interest, $S$, was the blood glucose level and the primary outcome of interest, $Y$, was low-density lipoprotein (LDL) cholesterol, both measured as change from baseline (randomization) to $4$ years post-randomization. In this study, our surrogate marker, blood glucose level, was routinely measured because it was necessary to monitor treatment. Crucially, blood glucose may serve as a valid surrogate for LDL because treatment-induced changes in glycemia are expected to propagate downstream to changes in lipid metabolism, meaning that a treatment's effect on LDL may be largely captured by its effect on glucose. Glycemic control, measured via blood glucose, contributes to lipid abnormalities and vascular damage in diabetes.\citep{jenkins2015biomarkers,brands2004cerebral,taskinen2003diabetic} Chronic hyperglycemia promotes insulin resistance, hepatic lipid overproduction, and systemic inflammation, all of which influence LDL levels and atherogenic processes.\citep{meshkani2009hepatic,semenkovich2006insulin} Our primary outcome, LDL cholesterol, contributes to plaque buildup in the arteries and is an important indicator of both microvascular complications, such as diabetic retinopathy (damage to the blood vessels in the retina of the eye) and diabetic nephropathy (kidney disease), and cardiovascular risk.\citep{jenkins2015biomarkers,trevisan2006lipids,goodman1988report} 

\begin{figure}[ht]
    \centering
    \caption{Distributions of the surrogate (change in glucose; top left panel) and the primary outcome (change in LDL cholesterol; top right panel) by treatment group, and the surrogate versus primary outcome (bottom panel) within each treatment group in the Diabetes Control and Complications Trial. The solid lines in the bottom panel are loess smoothers, and the shaded area represents the $95\%$ confidence intervals around them.}\label{dcct_fig}
    \includegraphics[width=1\linewidth]{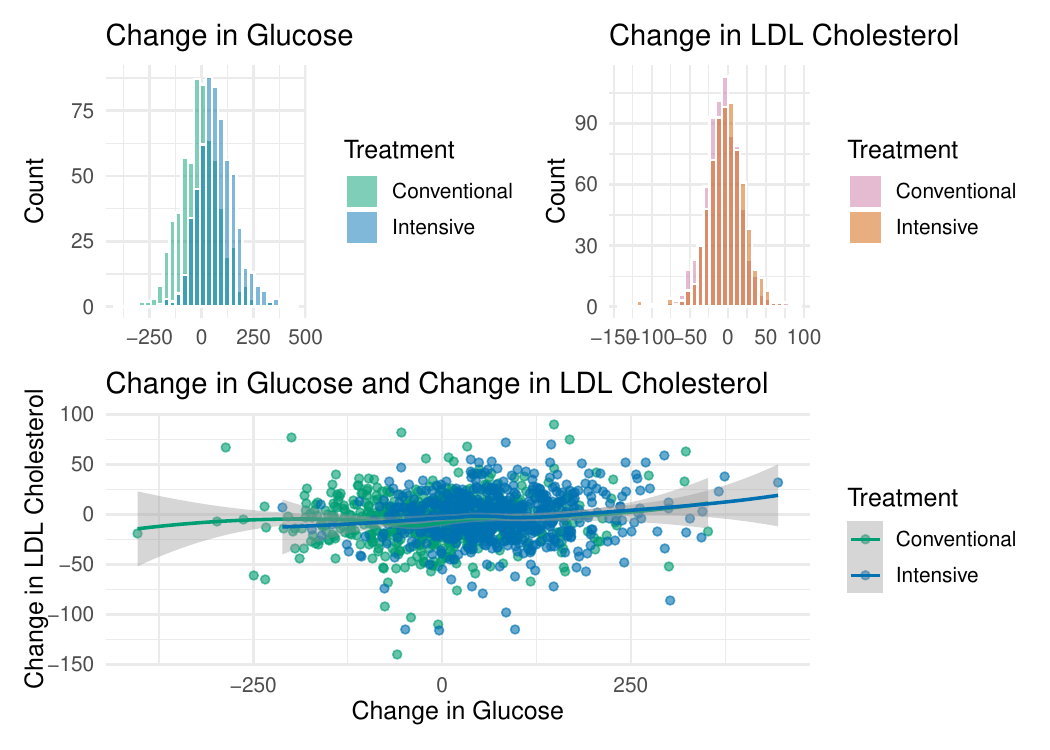}
\end{figure}

A total of $208$ out of $n = 1414$ ($14.7\%$) participants were missing the surrogate $S$, blood glucose, at the $4$ year measurement. Figure \ref{dcct_fig} shows the distributions of the surrogate and primary outcome by treatment group, as well as a scatterplot of the surrogate versus the primary outcome within each treatment group. The top left panel of the figure shows that the supports for the surrogate in each group do not completely overlap. Thus, we would expect the nonparametric approach, even without missingness, to possibly result in some bias in estimating the PTE. The bottom panel also shows a relatively linear relationship between the surrogate and the primary outcome within each group indicating that the parametric approach, which uses linear regression, may be reasonable. 

Full estimation results for the PTE estimate with quantile 95\% confidence intervals are shown in Table \ref{dcctresults}. Using a complete case analysis, the proportion of the treatment effect on LDL cholesterol explained by the treatment effect on blood glucose was $\widehat{R}_S = 0.659$ (95\% CI: $0.140$, $2.477$) using parametric estimation and $\widehat{R}_S = 0.856$ (95\% CI: $0.306$, $2.293$) using nonparametric estimation. The corresponding missingness-adjusted estimates were broadly similar to the complete case results. Although the true PTE is unknown in this application, this similarity between estimates with/without corrections suggests that bias due to missingness may be limited. (As discussed in the simulation studies, complete case estimates can be be unbiased when the surrogate marker is MCAR or MAR given treatment.) Among the parametric PTE estimators, the proposed SMLE approach yielded a $31\%$ narrower confidence interval than the IPW estimator, consistent with the efficiency gains observed in the simulation study. Thus, SMLE appeared to provide a principled correction for missingness while improving statistical efficiency relative to IPW.

The parametric and nonparametric PTE estimates differed meaningfully. However, given the limited overlap in the surrogate distributions across treatment groups in this dataset, the parametric model may have offered greater stability and reliability in this setting. Overall, the point estimates ($0.658$ to $0.856$) suggested that blood glucose explained a moderate proportion of the treatment effect on LDL cholesterol. Nevertheless, the lower bounds of the confidence intervals for all methods remained well below $0.5$, providing evidence that glucose cannot be considered as a valid surrogate marker for LDL cholesterol in this context.

\section{Discussion}
In this paper, we propose multiple methods that enable robust and efficient evaluation of surrogate markers in the presence of missing surrogate marker data, both within a parametric and nonparametric estimation framework. Our simulation results demonstrate good performance of the proposed methods, and our application to the DCCT study further illustrates their practical utility. To allow others to easily use the methods in practice, we have created an R package, \textit{MissSurrogate}, implementing them, which is available at \url{https://github.com/sarahlotspeich/missSurrogate}. 

It would be valuable to explore additional missing data approaches beyond IPW within the nonparametric PTE framework. Our proposed SMLE is specific to the parametric PTE approach and is not applicable in the nonparametric framework because no full likelihood is specified. Instead, a multiple imputation (MI) strategy (another model-based approach) could be considered, although it would generally require some degree of model specification. In the context of confidence interval construction under MI, prior work outside the surrogate marker setting suggests that combining bootstrap procedures with MI can substantially improve performance, particularly when concerns arise regarding uncongeniality or model misspecification.\citep{bartlett2020bootstrap}
The proposed SMLE could also be extended to condition on information beyond $Z$ in the nonparametric component, for example, through the use of B-spline sieves to smooth continuous variables on which the surrogate marker could depend.\citep{Tao2017} Moreover, while our focus has been on missing surrogate marker data, extending the proposed methods to accommodate missingness in the primary outcome would be an important direction for future research. Finally, our framework assumes that missingness is either missing completely at random or missing at random. In settings where data are missing not at random, the proposed methods would generally not be valid, and alternative strategies, like sensitivity analyses, would be required.

\section*{Funding}
This work was supported in part by NIDDK grant R01DK118354 (PI:Parast).

\section*{Acknowledgments}
The Diabetes Control and Complications Trial (DCCT) and its follow-up the Epidemiology of Diabetes Interventions and Complications (EDIC) study were conducted by the DCCT/EDIC Research Group and supported by National Institutes of Health (NIH) grants and contracts and by the General Clinical Research Center Program (GCRC), the National Center for Research Resources (NCRR). The resources from the DCCT/EDIC study were supplied by NIDDK Central Repository (NIDDK-CR). This manuscript was not prepared under the auspices of the DCCT/EDIC study and does not represent analyses or conclusions of the DCCT/EDIC study group, NIDDK-CR, or NIH.\vspace*{-8pt}

\section*{Supporting Information}

Additional supporting information may be found in the online version of the article at the publisher’s website. An R package \textit{missSurrogate} that implements the missing data methods described in this article is available at \url{https://github.com/sarahlotspeich/missSurrogate}. The R scripts and data needed to replicate the simulation studies from Section~\ref{sims} are available at \url{https://github.com/sarahlotspeich/evaluate_missing_surrogates}.

\section*{Data Availability Statement}
The DCCT study data used in this paper are publicly available upon request from the NIDDK Data Repository and completion of a data use agreement: https://repository.niddk.nih.gov/study/159

\bibliographystyle{unsrt}  
\bibliography{bib}

% Figure 1

% Figure 2

%TC:ignore 
% Figure 3
%TC:endignore 

%TC:ignore 
% Table 1
\begin{table}[ht]
\centering
%\resizebox{\columnwidth}{!}{
\begin{threeparttable}
\caption{Simulation results for $\widehat{R}_S$, the proportion of treatment effect explained (PTE) in Setting 1, where the surrogate $S$ was missing completely at random for $32$--$39\%$ of patients in the sample of $n = 2000$. With $R_S = 0.5$, the surrogate $S$ was of moderate strength.} \label{tab:mcar}
\centering
\begin{tabular}{rrrccccc}
\toprule
\textbf{Method} & \textbf{Bias} & \textbf{\% Bias} & \textbf{ESE} & \textbf{ASE} & \textbf{CP-N} & \textbf{CP-Q} & \textbf{RE}\\
\midrule
\addlinespace[0.3em]
\multicolumn{8}{l}{\textit{\textbf{PTE Estimator: Nonparametric}}}\\
\hspace{1em}Gold Standard & $-0.001$ & $-0.200$ & $0.021$ & $0.020$ & $0.952$ & $0.944$ & $1.000$\\
\hspace{1em}Complete Case & $-0.001$ & $-0.300$ & $0.025$ & $0.025$ & $0.955$ & $0.954$ & $0.685$\\
\hspace{1em}IPW & $-0.001$ & $-0.200$ & $0.024$ & $0.025$ & $0.953$ & $0.947$ & $0.724$\\
\addlinespace[0.3em]
\multicolumn{8}{l}{\textit{\textbf{PTE Estimator: Parametric}}}\\
\hspace{1em}Gold Standard & $ 0.000$ & $ 0.000$ & $0.020$ & $0.020$ & $0.950$ & $0.946$ & $1.000$\\
\hspace{1em}Complete Case & $ 0.000$ & $ 0.000$ & $0.024$ & $0.024$ & $0.943$ & $0.946$ & $0.677$\\
\hspace{1em}IPW & $ 0.000$ & $ 0.000$ & $0.024$ & $0.024$ & $0.943$ & $0.946$ & $0.677$\\
\hspace{1em}SMLE & $ 0.000$ & $ 0.000$ & $0.022$ & $0.022$ & $0.940$ & $0.941$ & $0.793$\\
\bottomrule
\end{tabular}
\begin{tablenotes}[flushleft]
\item{\em Note:} \simtabnote
\end{tablenotes}
\end{threeparttable}
%}
\end{table}
%TC:endignore 

%TC:ignore 
% Table 2
\begin{table}[ht]
\centering
%\resizebox{\columnwidth}{!}{
\begin{threeparttable}
\centering
\caption{Simulation results for $\widehat{R}_S$, the proportion of treatment effect explained (PTE) in Settings $2$ and $3$, when the surrogate marker $S$ was missing at random depending on treatment $Z$ (top portion) and depending on the primary outcome $Y$ (bottom portion) in the sample of $n = 2000$. With $R_S = 0.5$, the surrogate $S$ was of moderate strength.}\label{tab:mar}
\begin{tabular}{rrrccccc}
\toprule
\textbf{Method} & \textbf{Bias} & \textbf{\% Bias} & \textbf{ESE} & \textbf{ASE} & \textbf{CP-N} & \textbf{CP-Quant} & \textbf{RE}\\
\midrule
\addlinespace[0.3em]
\multicolumn{8}{c}{\textit{\textbf{Missingness Depends on Treatment $\pmb{Z}$}}}\\
\addlinespace[0.3em]
\multicolumn{8}{l}{\textit{PTE Estimator: Nonparametric}}\\
\hspace{1em}Gold Standard & $-0.002$ & $-0.400$ & $0.020$ & $0.021$ & $0.953$ & $0.943$ & $1.000$\\
\hspace{1em}Complete Case & $-0.002$ & $-0.500$ & $0.025$ & $0.026$ & $0.954$ & $0.950$ & $0.639$\\
\hspace{1em}IPW & $-0.002$ & $-0.500$ & $0.025$ & $0.026$ & $0.955$ & $0.947$ & $0.652$\\
\addlinespace[0.3em]
\multicolumn{8}{l}{\textit{PTE Estimator: Parametric}}\\
\addlinespace[0.3em]
\hspace{1em}Gold Standard & $-0.001$ & $-0.200$ & $0.020$ & $0.020$ & $0.941$ & $0.941$ & $1.000$\\
\hspace{1em}Complete Case & $-0.001$ & $-0.200$ & $0.025$ & $0.025$ & $0.954$ & $0.952$ & $0.640$\\
\hspace{1em}IPW & $-0.001$ & $-0.200$ & $0.025$ & $0.025$ & $0.953$ & $0.948$ & $0.640$\\
\hspace{1em}SMLE & $ 0.000$ & $ 0.000$ & $0.023$ & $0.022$ & $0.948$ & $0.943$ & $0.755$\\
\addlinespace[0.3em]
\multicolumn{8}{c}{\textit{\textbf{Missingness Depends on Primary Outcome $\pmb{Y}$}}}\\
\addlinespace[0.3em]
\multicolumn{8}{l}{\textit{PTE Estimator: Nonparametric}}\\
\addlinespace[0.3em]
\hspace{1em}Gold Standard & $-0.002$ & $-0.400$ & $0.020$ & $0.021$ & $0.957$ & $0.950$ & $1.000$\\
\hspace{1em}Complete Case & $ 0.019$ & $ 3.900$ & $0.024$ & $0.024$ & $0.857$ & $0.876$ & $0.729$\\
\hspace{1em}IPW & $-0.003$ & $-0.700$ & $0.025$ & $0.025$ & $0.945$ & $0.942$ & $0.661$\\
\addlinespace[0.3em]
\multicolumn{8}{l}{\textit{PTE Estimator: Parametric}}\\
\addlinespace[0.3em]
\hspace{1em}Gold Standard & $-0.001$ & $-0.100$ & $0.020$ & $0.020$ & $0.946$ & $0.946$ & $1.000$\\
\hspace{1em}Complete Case & $ 0.021$ & $ 4.300$ & $0.023$ & $0.023$ & $0.820$ & $0.844$ & $0.729$\\
\hspace{1em}IPW & $-0.001$ & $-0.300$ & $0.025$ & $0.024$ & $0.948$ & $0.939$ & $0.648$\\
\hspace{1em}SMLE & $ 0.002$ & $ 0.500$ & $0.023$ & $0.022$ & $0.936$ & $0.943$ & $0.750$\\
\bottomrule
\end{tabular}
\begin{tablenotes}[flushleft]
\item{\em Note:} \simtabnote
\end{tablenotes}
\end{threeparttable}
%}
\end{table}
%TC:endignore 

%TC:ignore 
% Table 3
\begin{table}[ht]
\centering
%\resizebox{\columnwidth}{!}{
\begin{threeparttable}
\centering
\caption{Simulation results for $\widehat{R}_S$, the proportion of treatment effect explained (PTE) in Setting $5$, when the distributions of the surrogate markers $S$ do not overlap between the treatment and control groups.}\label{tab:overlap}
\begin{tabular}{rrrccccc}
\toprule
\textbf{Method} & \textbf{Bias} & \textbf{\% Bias} & \textbf{ESE} & \textbf{ASE} & \textbf{CP-N} & \textbf{CP-Quant} & \textbf{RE}\\
\midrule
\addlinespace[0.3em]
\multicolumn{8}{l}{\textit{PTE Estimator: Nonparametric}}\\
\hspace{1em}Gold Standard & $-0.084$ & $-16.800$ & $0.083$ & $0.059$ & $0.569$ & $0.481$ & $1.000$\\
\hspace{1em}Complete Case & $-0.098$ & $-19.500$ & $0.088$ & $0.062$ & $0.544$ & $0.438$ & $0.893$\\
\hspace{1em}IPW & $-0.096$ & $-19.300$ & $0.090$ & $0.064$ & $0.555$ & $0.456$ & $0.851$\\
\addlinespace[0.3em]
\multicolumn{8}{l}{\textit{PTE Estimator: Parametric}}\\
\hspace{1em}Gold Standard & $ 0.001$ & $  0.200$ & $0.020$ & $0.020$ & $0.946$ & $0.942$ & $1.000$\\
\hspace{1em}Complete Case & $-0.006$ & $ -1.100$ & $0.025$ & $0.025$ & $0.945$ & $0.943$ & $0.641$\\
\hspace{1em}IPW & $ 0.001$ & $  0.200$ & $0.024$ & $0.025$ & $0.953$ & $0.954$ & $0.654$\\
\hspace{1em}SMLE & $ 0.002$ & $  0.500$ & $0.025$ & $0.025$ & $0.949$ & $0.950$ & $0.636$\\
\bottomrule
\end{tabular}
\begin{tablenotes}[flushleft]
\item{\em Note:} \simtabnote
\end{tablenotes}
\end{threeparttable}
%}
\end{table}
%TC:endignore 

%TC:ignore 
%Table 4
\begin{table}[ht]
%\resizebox{\columnwidth}{!}{
\centering
\begin{threeparttable}
\caption{Results applying proposed methods to evaluate blood glucose level as a surrogate marker in the Diabetes Control and Complications Trial using the proportion of treatment effect explained (PTE). \label{dcctresults}}
\begin{tabular}{rccc}
\toprule
\textbf{Method} & \textbf{$\pmb{R_S}$ Estimate} & \textbf{$\pmb{95\%}$ CI for $\pmb{R_S}$} & \textbf{CI Width}\\
\midrule
\addlinespace[0.3em]
\multicolumn{4}{l}{\textit{\textbf{PTE Estimator: Nonparametric}}}\\
\hspace{1em}Complete Case & $0.856$ & $(0.306, 2.293)$ & $1.987$ \\
\hspace{1em}IPW & $0.842$ & $(0.207, 2.485)$ & $2.278$ \\
\addlinespace[0.3em]
\multicolumn{4}{l}{\textit{\textbf{PTE Estimator: Parametric}}}\\
\hspace{1em}Complete Case & $0.659$ & $(0.140, 2.477)$  & $2.337$ \\
\hspace{1em}IPW & $0.663$ & $(0.097, 2.912)$ & $2.815$ \\
\hspace{1em}SMLE & $0.658$ & $(0.131, 2.065)$ & $1.934$ \\
\bottomrule
\end{tabular}
\begin{tablenotes}[flushleft]
\item{\em Note:} \textbf{95\% CI} is the $95\%$ confidence interval for $R_S$, \textbf{CI Width} is the width of the quantile $95\%$ confidence interval for $R_S$ (calculated as the lower bound minus the upper bound). Abbreviations for the methods: inverse probability weighting (IPW) and semiparametric maximum likelihood estimation (SMLE).
\end{tablenotes}
\end{threeparttable}
%}
\end{table}
%TC:endignore 

\label{lastpage}

\end{document}